\documentclass[%
 reprint,
nofootinbib,
 amsmath,amssymb,
 aps,
]{revtex4-1}

\usepackage[a4paper, total={7.2in, 10in}]{geometry}

\usepackage{dcolumn}
\usepackage{bm}
\usepackage{natbib}
\usepackage{mathrsfs}
\usepackage[pdftex]{graphicx}
\usepackage{ifpdf}

\ifpdf 
  \usepackage[pdftex]{graphicx}
  \DeclareGraphicsExtensions{.pdf,.png,.jpg,.jpeg,.mps}
  \usepackage{pgf}
  \usepackage{tikz}
\else 
  \usepackage{graphicx}
  \DeclareGraphicsExtensions{.eps,.bmp}
  \usepackage{pgf}
  \usepackage{tikz}
  \usepackage{pstricks}
\fi
\usepackage{pdfpages}
\usepackage{pgffor}

\begin{document}


\title{Large Brillouin Amplification in Silicon}

\author{Eric A. Kittlaus$^{1,*}$, Heedeuk Shin$^{2,*}$, and Peter T. Rakich$^1$}
\makeatother

\affiliation{$^1$Department of Applied Physics, Yale University, New Haven, CT 06520 USA.\\
$^2$Department of Physics, Pohang University of Science and Technology, Pohang 37673, Korea.}

\date{\today}

\begin{abstract}

Strong Brillouin coupling has only recently been realized in silicon using a new class of optomechanical waveguides that yield both optical and phononic confinement. Despite these major advances, appreciable Brillouin amplification has yet to be observed in silicon. Using a new membrane-suspended silicon waveguide we report large Brillouin amplification for the first time, reaching levels greater than 5 dB for modest pump powers, and demonstrate a record low (5 mW) threshold for net amplification. This work represents a crucial advance necessary to realize high-performance Brillouin lasers and amplifiers in silicon.
\end{abstract}

\maketitle

Both\let\thefootnote\relax\footnotetext{*These authors contributed equally to this work.} Kerr and Raman nonlinearities are radically enhanced by tight optical-mode confinement in nanoscale silicon waveguides \protect{\cite{rong2004,rong2005,jalali2006,foster2006}}. Counter-intuitively, Brillouin nonlinearities are exceedingly weak in these same nonlinear waveguides \protect{\cite{rakichprx}}. Only recently have strong Brillouin interactions been realized in a new class of optomechanical structures that control the interaction between guided photons and phonons \protect{\cite{rakichprx,shinnatcomm,roel}}. With careful design, such Brillouin nonlinearities overtake all other nonlinear processes in silicon \protect{\cite{shinnatcomm,roel}}; these same Brillouin interactions are remarkably tailorable, permitting a range of hybrid photonic-phononic signal processing operations that have no analog in all-optical signal processing \protect{\cite{Vidal07,Li2008,zhang12,pant2014,shinpper}}.  Using this physics, the rapidly growing field of silicon-based Brillouin-photonics has produced new frequency agile RF-photonic notch filters \protect{\cite{Vidal07,zhang12,Li2013,marpaung}} and multi-pole bandpass filters \protect{\cite{shinpper}} as the basis for radio-frequency photonic (RF-photonic) signal processing. Beyond these specific examples, the potential impact of such Brillouin interactions is immense; frequency combs \protect{\cite{kang,braje09,Li2013}}, ultra-low phase-noise lasers \protect{\cite{olsson,Abedin12,Li14}}, sensors \protect{\cite{Li2008,Gao13,shinpper}}, optical isolation\protect{\cite{Yu2009,Kang2011,Huang11,kim2015}}, and an array of signal processing technologies \protect{\cite{Yao98,gaeta,Vidal07,loayssa,shinpper,marpaung,Li2013}} may be possible in silicon with further progress.

However, strong Brillouin amplification--essential to many new Brillouin-based technologies--has yet to be realized in silicon photonics.  Despite the creation of strong Brillouin nonlinearities in a range of new structures \protect{\cite{shinnatcomm,roel}}, nonlinear losses and free carrier effects have stifled attempts to demonstrate net optical amplification. Only recently, Van Laer \emph{et al.} reported 0.5 dB  (12\%) amplification \protect{\cite{van2015net}} using suspended silicon nanowire structures. Even with superb dimensional control, amplification diminishes with longer interaction lengths \protect{\cite{van2015net}}, highlighting the problem of dimensionally induced inhomogenous broadening \protect{\cite{wolff2015brillouin}}. Careful theoretical analyses by Wolff \emph{et al.}, suggest that large net amplification is fundamentally challenging to achieve in silicon nanowires at near-IR wavelengths due to nonlinear absorption \protect{\cite{wolffarxiv}}.

%

\begin{figure}[b]
\centering\vspace{-10pt}
\includegraphics[width=\linewidth]{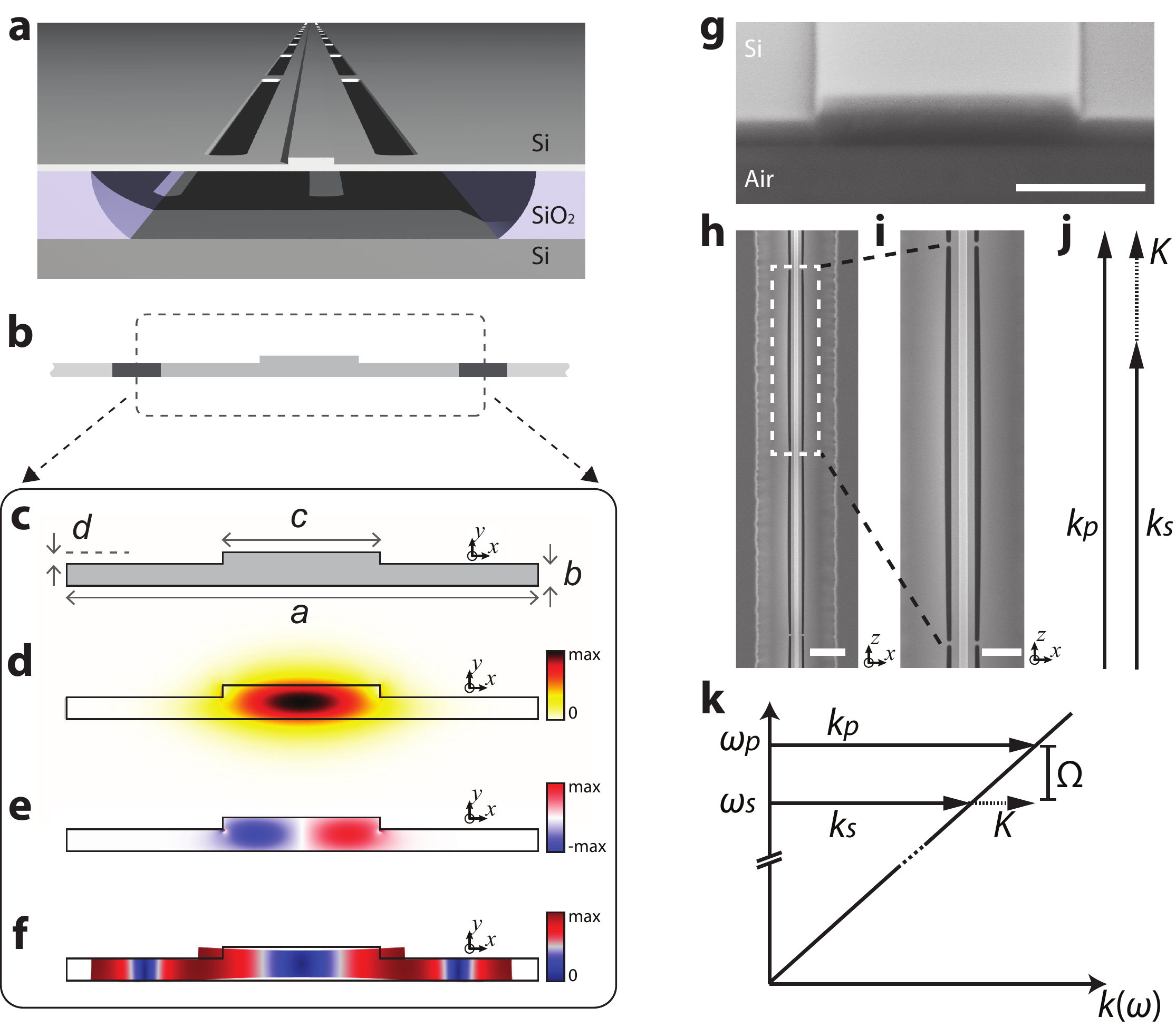}
\caption{Hybrid photonic-phononic silicon waveguide. (a) schematic of continuously suspended silicon Brillouin-active waveguide and (b) cross-section of the active region. (c) diagram showing critical device dimensions. (d), (e), and (f) are $E_x$ field of guided optical mode, $x$-component of electrostrictive body force, and the elastic displacement field of the Brillouin active phonon mode, respectively.  (g) cross-sectional SEM of waveguide core, while (h) and (i) show top-down SEM images of fabricated device. Scale bars indicate 500 nm, 10 $\mu$m and 5 $\mu$m in (g), (h), and (i) respectively. Phase-matching diagram of (j) is shown atop optical dispersion relation in (k).}
\label{fig:device}
\end{figure}

In this paper, we report large Brillouin amplification in silicon through an alternative device paradigm; using a new all-silicon membrane structure (Fig. \ref{fig:device}) that permits independent design of photonic and phononic modes, we demonstrate net amplification at remarkably low ($<\!5\!$ mW) pump powers, and record-high Brillouin amplification ($5.2\!$ dB) at 60 mW powers. These results represent a 30-fold improvement in net amplification over prior systems \protect{\cite{van2015net}}. We show that independent photonic and phononic control enables ultra-low ($<\! 0.2\,$ dB cm$^{-1}$) propagation losses, strong Brillouin coupling ($G_B\! >\! 10^3 \, $W$^{-1} $m$^{-1}$), and enhanced robustness to dimensional variations; this combination of properties yield dramatically improved performance. In contrast to Ref. \protect{\cite{shinnatcomm}}, this \emph{all-silicon} system permits dramatically reduced losses and 40-times larger nonlinear couplings. Using such strong nonlinearities, we demonstrate a unique form of cascaded-Brillouin energy transfer previously only observed in highly nonlinear micro-structured fibers \protect{\cite{kang}}. Hence, in addition to new Brillouin laser technologies \protect{\cite{kang,braje09,Li2013}}, these strong Brillouin couplings open the door for nonreciprocal signal processing schemes and comb generation \protect{\cite{kang,Yu2009,Kang2011,Huang11}} in silicon.

The Brillouin-active waveguide, of the type seen in Fig. \ref{fig:device}a, is continuously suspended over centimeter-lengths by a series of nanoscale tethers. The active region of the suspended waveguide (Fig. \ref{fig:device}b) is diagrammed in Fig. \ref{fig:device}c. Light is confined to the central ridge structure through total internal reflection, which supports guidance of the low-loss TE-like guided optical mode at $\lambda=1550$ nm (Fig. \ref{fig:device}d). Figure \ref{fig:device}e shows the electrostrictive optical force distribution generated by this mode; these optical forces mediate efficient coupling to the guided phonon mode (Fig. \ref{fig:device}f) at GHz frequencies. Confinement of this guided phonon mode is produced by the large acoustic impedance mismatch between silicon and air. Since the optical mode is confined to the central ridge, of width $c$, and the phonon mode extends throughout the membrane, of width $a$, the photon and phonon modes can be tailored independently. This independent control is used to minimize sensitivity to sidewall roughness while simultaneously maximizing photon-phonon coupling. 



Co-propagating pump and Stokes waves of frequencies $\omega_p$ and $\omega_s$ are guided in the same TE-like optical mode and couple through parametrically generated acoustic phonons with frequency $\Omega_B = \omega_p-\omega_s$. Coupling is mediated by guided phonon modes that satisfy the phase matching condition $K(\Omega_B) = k(\omega_p)-k(\omega_s)$ sketched in Fig 1j-k, where $K(\Omega)$ and $k(\omega)$ are the acoustic and optical dispersion relations. In the forward scattering case, phase matching requires $K(\Omega_B) \cong (\Omega_B/v_g)$, where $v_g \equiv (\partial \omega/\partial k)$ is the group velocity of the optical mode. Hence, phonons that mediate forward-SBS have a vanishing longitudinal wave-vector. As described in Ref. \protect{\cite{shinnatcomm}, a set of guided acoustic waves (or Lamb-waves) with exceedingly low ($<\!\! 1\!$ m/s) group velocities satisfy this condition. The underlying dynamics of this process are similar to that first observed in optical fibers \protect{\cite{kang}}, permitting both signal amplification and cascaded energy transfer that is distinct in nature from the more widely studied backwards-SBS process \protect{\cite{agrawal2007nonlinear}}.  

These devices are fabricated from a crystalline silicon layer through a silicon-on-insulator (SOI) fabrication process (see Methods). Scanning electron micrographs of the device cross-section and top-down view are shown in Fig. 1g and Fig. 1h-i respectively. The device consists of 80 nm $\times$ 1 $\mu$m wide ridge that sits atop a 3 $\mu$m wide, 160 nm thick silicon membrane. Each suspended region (seen in Fig. 1i) is supported by symmetrically placed nanoscale tethers spaced every 50 $\mu$m along the waveguide length; this design enables robust fabrication of several centimeter long Brillouin-active waveguides. Note that the supporting tethers have negligible contribution to the optical losses; linear propagation losses were measured to be $0.18\pm0.02$ dB$\,$cm$^{-1}$ through cutback measurements. (See Supplementary Section S2.) Such ultra-low losses contrasts sharply with Ref. \protect{\cite{shinnatcomm}}, where N-H bonds in the SiN layer produced excess absorption. In what follows, we examine Brillouin interactions in a 2.9 cm long continuously suspended Brillouin active device.

%

\begin{figure}[t]
\centering
\includegraphics[width=\linewidth]{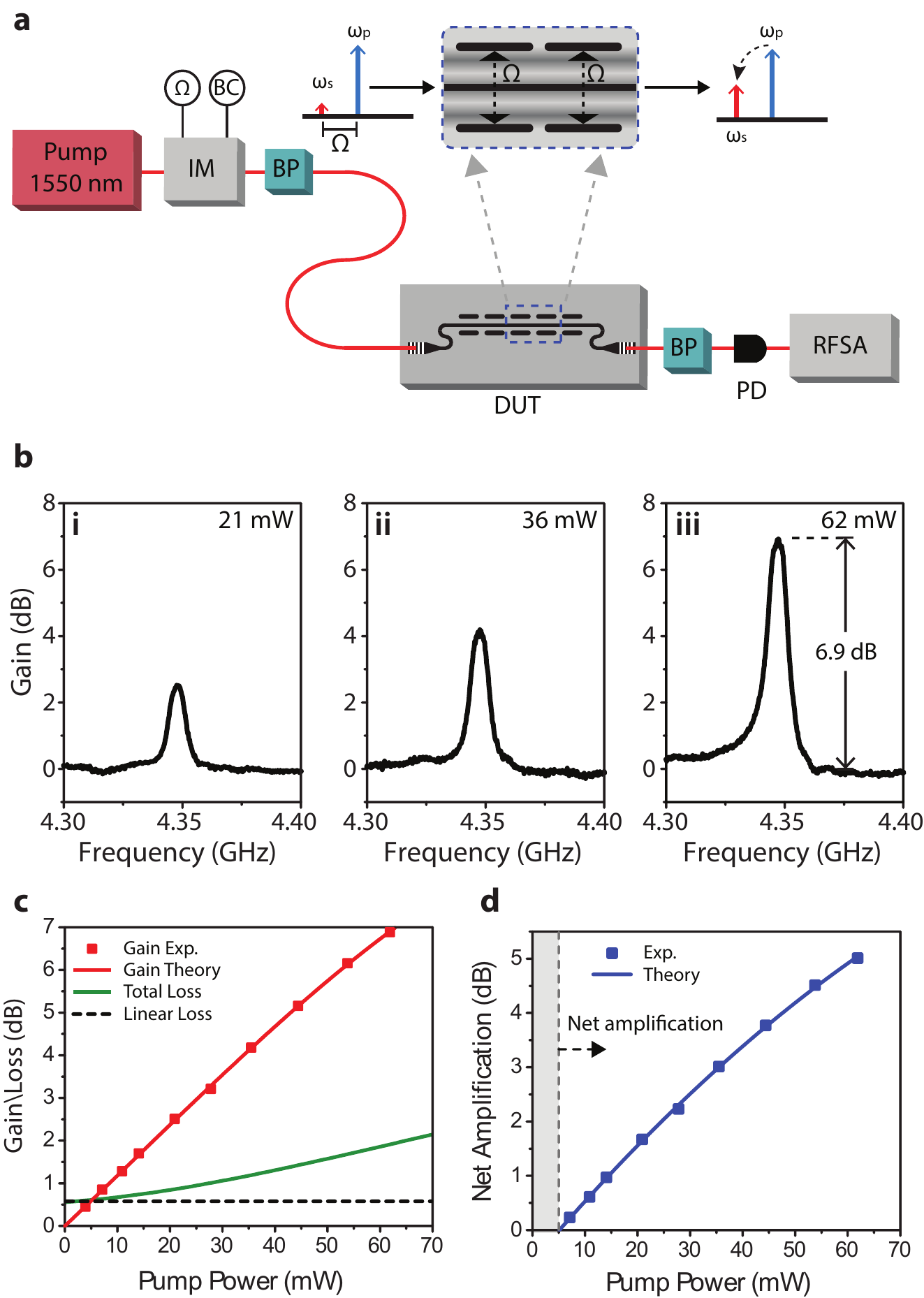}
\caption{Experimental results showing Brillouin gain and net amplification. (a) Diagram of the experimental apparatus; abbreviations detailed in Methods. Panels i, ii, and iii of (b) are the Brillouin gain spectra obtained for powers 21, 36, and 62 mW respectively. (c) plots peak gain (red), linear loss (dash) and total loss (green) versus pump power at 1550 nm wavelengths. (d) Net amplification as a function of pump power. The threshold for amplification is 5 mW.}
\label{fig:gain}
\end{figure}

Direct measurements of the Brillouin gain were performed using the apparatus of Fig. \ref{fig:gain}a. Bright pump- and weak probe-beams are synthesized from the same continuous-wave laser using an intensity modulator and band-pass filter. At the device output, the magnitude of probe signal is measured as an RF beat-note. The probe-beam is swept through the Brillouin resonance by varying the modulation frequency, permitting measurement of the Brillouin gain spectrum. Figures \ref{fig:gain}b.i-iii show the forward-Brillouin gain spectra for pump-wave powers of 21, 36, and 62 mW respectively. These spectra reveal a high quality-factor ($Q = 680$) Brillouin resonance at 4.35 GHz, demonstrating remarkable robustness to dimensional variations. Figure 2e shows the peak Brillouin gain as a function of pump power, reaching a maximum value of 6.9 dB at the highest (62 mW) pump power, consistent with a Brillouin gain coefficient of $G_B = 1152  \pm  54$ W$^{-1}$m$^{-1}$.  Independent measurements performed through heterodyne four-wave mixing spectroscopy yield a value of $G_B = 1120\,  \pm \, 180 $ W$^{-1}$m$^{-1}.$  (See Supplementary Section S3) Both measurements show good agreement with the predicted Brillouin frequency ($4.41$ GHz) and Brillouin gain ($1020 \pm  190$ W$^{-1}$m$^{-1}$) obtained from finite element simulations. 



The net Brillouin amplification (Fig. 2f) is obtained by subtracting the total loss (green) from the peak Brillouin gain (red) in Fig. 2c. These data reveal a peak on-chip amplification of 5.2 dB at 62 mW powers; moreover, the threshold for net amplification occurs at record-low ($<5\,$mW) optical powers owing to the low propagation losses and large Brillouin gain coefficient of this system. Note that the total waveguide loss represents the sum of the measured linear and nonlinear waveguide losses at a given power. (See Supplementary Section S2.) Through these measurements, net amplification (5.2 dB) was limited only by the power handling of the tapered input couplers; amplification continues to grow at the highest pump powers (62 mW). Hence, further amplification is clearly achievable with improved input coupler designs. 







\begin{figure}[b]
\centering
\includegraphics[width=2.99in]{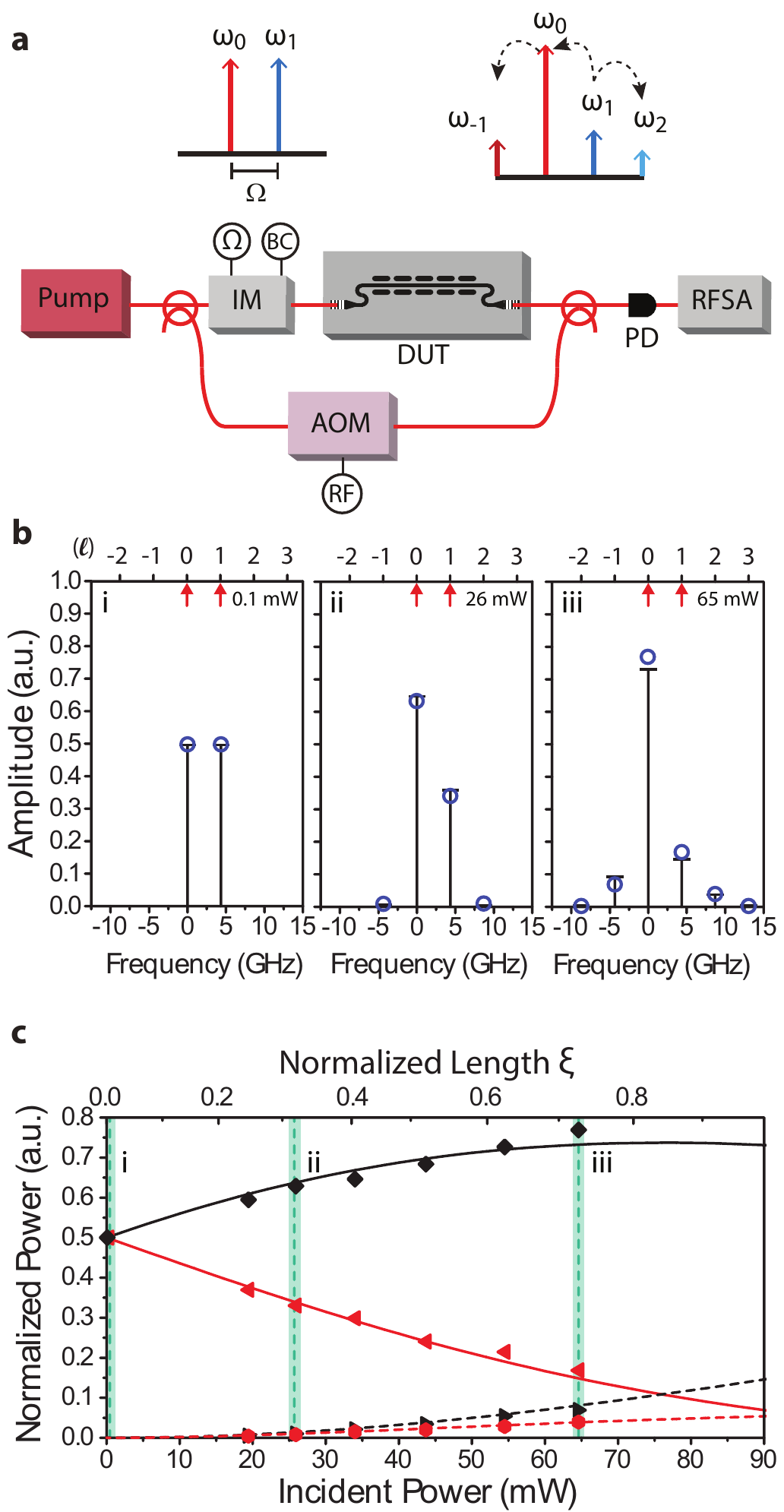}
\caption{Setup and results of an energy transfer (two-tone) experiment (a) experimental diagram to measure power transfer driven by two pump fields of equal magnitude (b) theory (black) and data (blue circles) representing the frequency spectrum of output light for total input powers of 0.1, 26, and 65 mW. (c) power transfer as a function of on-chip power showing theoretical calculations (lines) and measured amplitudes for pump fields $\omega_0$ and $\omega_1$ (solid black, solid red) and first cascaded fields $\omega_{-1}$ and $\omega_{2}$ (solid red, dashed red) fields. The vertical blue lines correspond to panels i-iii above.}
\label{fig:etransfer}
\end{figure}

Such large nonlinear couplings also enable cascaded forward-Brillouin energy transfer. This interaction, previously only observed in highly nonlinear micro-structured fibers \protect{\cite{kang}}, is quantified by injecting two drive fields at frequencies $\omega_0$ and $\omega_1$ into the Brillouin-active waveguide using the apparatus in Fig. \ref{fig:etransfer}a. Nonlinear coupling between these equal intensity drive fields, with frequency separation $\Omega_B$, produce resonant phonon-mediated energy transfer (sketched in the top panel of Fig. \ref{fig:etransfer}a). Heterodyne spectral analysis is used to quantify cascaded energy transfer to successive Stokes and anti-Stokes orders, of frequencies $\{\omega_l\}$, at the device output; three characteristic spectra are shown in Fig. \ref{fig:etransfer}b-d, corresponding to injected powers $P = P_0 =P_1$ of 0.1 mW, 13 mW and 32 mW respectively. These data reveal 62\% power transfer from $l\!=\!1$ to $l\!=\!\{-2,-1,0,+2,+3\}$ orders through strong light driven acousto-optic energy transfer.  


This cascading process is unique to forward Brillouin scattering; through this process the cascaded optical fields coherently drive the same phonon field producing successive parametric frequency shifts \protect{\cite{kang}}. The relevant figure of merit for cascaded energy transfer is the power-gain-length product $\xi$, which is defined as $\xi \equiv G_B \sqrt{(P_0\,P_1)} L_{\textup{eff}} = G_B P L_{\textup{eff}}$. Here, $L_{\textup{eff}}$ is the effective interaction length taking into account nonlinear loss (see Supplementary Section S2). At the highest tested power, this device admits a normalized propagation length $\xi \!= \!0.71$. As seen from both experiment and theory in Fig. 3c, this system approaches maximal depletion of the $\omega_1$ drive-field, resulting in significant energy transfer to higher order Stokes and anti-Stokes orders. This process is a basis for wideband nonreciprocal energy transfer, comb generation, and waveform synthesis \protect{\cite{kang}}.

More generally, new silicon-based Brillouin laser and amplifier technologies hinge upon the relative strength of Brillouin coupling and linear/nonlinear losses. In this regard, the planar waveguide topology offers several advantages over nanowire systems. Greatly reduced sensitivity of this optical mode to lithographic sidewall roughness permits ultra-low ($0.18 \pm 0.02$ dB cm$^{-1}$) propagation losses \protect{\cite{barwicz2005three,dong2010low}}, enabling interaction lengths of 25 cm, more than an order of magnitude larger than prior systems \protect{\cite{shinnatcomm,roel,van2015net}}. In addition, this same waveguide geometry drastically shortens the effective free-carrier lifetimes ($\sim2$ ns) relative to bulk ($>10$ ns) through rapid in-plane diffusion of carriers \protect{\cite{clapsraman}}. (See Supplementary Sections S3-4.) 
Short free carrier lifetimes, combined with a reduced two-photon absorption coefficient, result in $15$-times lower nonlinear losses than the nanowire systems analyzed in Ref. \protect{\cite{wolffarxiv}}. Together these properties yield an unprecedented Brillouin gain figure of merit of $\mathscr{F}$ = 5.2 \protect{\cite{wolffarxiv}}. Hence, this unique hybrid photonic-phononic waveguide design represents a significant advance over prior silicon nanowire waveguide systems. Moreover, this improved figure of merit suggests significant opportunity for high-efficiency Brillouin amplification and cascaded energy transfer \protect{\cite{wolffarxiv}}. 

Thus far, we have focused on individual Brillouin-active waveguides of 2.9 cm length, and the injected optical powers have been limited to $\sim 60$ mW by the input coupling method. With high power input couplers, this waveguide geometry readily supports guided powers of 150 mW \protect{\cite{shinpper}}. Furthermore, low propagation losses permit exceptionally large (25 cm) propagation lengths. Taking nonlinear losses into account, these conditions would enable 30 dB of amplification over a 20 cm waveguide length; a cascaded energy transfer figure of merit of $\xi \cong 3.8$ is also achieved under these conditions, corresponding to efficient energy transfer to more than 20 comb lines. However, this level of performance is only possible if inhomogeneous broadending does not diminish the Brillouin gain for increasing device lengths \protect{\cite{shinnatcomm,roel,van2015net}}.

To explore the impact of inhomogeneous broadening on Brillouin gain, we fabricated devices with lengths ranging from 500 microns to 2.9 cm. (See Supplementary Section S7.) The shortest waveguide lengths reveal gains of $G_B = 1840\,  \pm \, 130 $ W$^{-1}$m$^{-1}$; this increased gain is consistent with measured higher effective phononic Q-factors ($Q \cong 1019 \,  \pm \, 59 $). As lengths are increased (L = 0.5, 2.5,  5 and 29 mm) the Q-factor and gain monatonically decrease in a manner consistent with inhomogeneous broadening. Note however, the majority of dimensionally induced broadening occurs over the first 5 milimeters. Only a marginal (15\%) increase in linewidth is seen for a 6-fold increase in length (i.e., from 0.5 cm to 2.9 cm), demonstrating a greatly improved net Brillouin amplification with increasing device length.


These results contrast sharply with recent studies of nanowire waveguides \protect{\cite{shinnatcomm,roel,van2015net}}, where inhomogeneous broadening becomes problematic as device dimensions are increased. In the membrane system, the Brillouin-active phonon mode is much more robust to dimensional variations, highlighting another advantage of independent photonic and phononic design. Expanding on this point, we analyze the sensitivity of Brillouin frequency with dimensional variations. FEM simulations reveal resonant frequency changes of $1.4 \,$MHz$\,$nm$^{-1}$, $0.8\,$MHz$\,$nm$^{-1}$, and $0.7 \,$MHz$\,$nm$^{-1}$ with variations in device dimensions $a,$ $b,$ and $c$, as diagrammed in Fig. 1c. (See Supplementary Section S7.) 
This degree of dimensional sensitivity is 10 times smaller than that of nanowire systems ($19\,$MHz$\,$nm$^{-1}$)  \protect{\cite{roel}}, which explains the improved robustness and scalability of this system.


To explore the potential for higher single-pass amplification, we also fabricated Brillouin-active waveguide structures with lengths up to 8.7 cm. These  structures produced reliable low-loss optical transmission. However, for lengths greater than 2.9 cm it was necessary to wrap the waveguide using a serpentine device geometry. Even for short waveguide lengths, this change in layout also altered the character of the resonance lineshape, suggesting that systematic  effects (e.g., proximity effect, stress propagation, and process variations) likely play a role in excess broadening observed in these modified device designs. Hence, to approach the theoretical limits of Brillouin gain and nonlinear coupling described above, it may be necessary to implement systematic frequency compensation across fabricated devices.  

While further work is required to approach the theoretical maximum of Brillouin amplification, we have shown that there is immense potential for large Brillouin amplification in silicon at near-IR wavelengths. The demonstrated Brillouin amplification (and improved figures of merit) already permits high performance Brillouin laser and signal processing technologies. For instance, the ultra low propagation losses of this waveguide geometry translate to optical cavity Q-factors of $> 4$ million. Using this new waveguide design, efficient and tailorable all-silicon Brillouin lasers--with thresholds approaching $10$ mW--are now within reach.   

In conclusion, using a new Brillouin-active waveguide system, we have demonstrated large Brillouin amplification necessary to achieve high-performance on-chip Brillouin lasers and signal processing technologies. Through independent photonic and phononic control, we have shown that a combination of ultra-low propagation losses, strong Brillouin coupling, and robustness to dimensional variations yields dramatic improvements in Brillouin performance over prior systems. Using this system, we demonstrated net amplification of over 5 dB  for modest (60 mW) pump powers, and significant cascaded energy transfer. Beyond this device study, we have shown that there is in fact immense potential for flexible silicon-based Brillouin photonics at near-IR wavelengths. This combination of high nonlinearity with robust optical and acoustic performance opens the door for a wide range of hybrid silicon photonic-phononic technologies for RF and photonic signal processing \protect{\cite{Yao98,Vidal07,loayssa,shinpper,marpaung,Li2013}}, waveform and pulse synthesis \protect{\cite{kang,braje09,Li2013}}, optical isolators and filtering, and new light sources \protect{\cite{Yu2009,Kang2011,Huang11}}.




\subsubsection{Fabrication Methods}
The silicon waveguides were written on a silicon-on-insulator chip with a 3 $\mu$m oxide layer using electron beam lithography on hydrogen silsesquioxane photoresist. Following development, a Cl$_2$ reactive ion etch was employed to etch the ridge waveguide structure. After a solvent cleaning step, slots were written to expose the oxide layer, again with electron beam lithography of ZEP520A photoresist and Cl$_2$ RIE. The device was then wet released via a 49\% hydrofluoric acid etch of the oxide undercladding. The waveguide structure under test is comprised of 570 suspended segments.


\subsubsection{Experimental Methods}
Both experiments used a pump laser operating around 1550 nm. Light is coupled in and out of the waveguide through the use of grating couplers with measured coupling losses of 6.5 dB. The following abbreviations are used in the experimental diagrams: 

Fig. 2a: IM Mach-Zehnder intensity modulator, BC DC bias controller, BP band-pass filter, DUT device under test, PD photodetector, RFSA radio frequency spectrum analyzer.  

Fig. 3a: IM Mach-Zehnder intensity modulator, BC DC bias controller, AOM acousto-optic modulator, DUT device under test, PD photodetector, RFSA radio frequency spectrum analyzer.  

\subsubsection{Acknowledgements}
This work was supported by the MesoDynamic Architectures program at DARPA under the direction of Dr. Daniel Green. We thank Prashanta Kharel for technical discussions involving phononic systems and nonlinear interactions, and Dr. Michael Rooks and Dr. Michael Power for assistance with process development. We are grateful to Dr. Ryan Behunin, Dr. William Renninger, and Dr. Whitney Purvis Rakich for careful reading and critique of this manuscript.

\subsubsection{Author Contributions}
E.A.K. and H.S. fabricated the waveguide devices. P.R., H.S., and E.A.K. developed multi-physics simulations for and designed the devices. E.A.K. and H.S. conducted experiments with the assistance of P.R. P.R., H.S., and E.A.K. developed analytical models to interpret measured data. All authors contributed to the writing of this paper.

\subsubsection{Additional Information}
\noindent \textbf{Competing financial interests:} The authors declare no competing financial interests.

\bibliographystyle{apsrev4-1} 
\bibliography{cites}

\begin{thebibliography}{34}%
\makeatletter
\providecommand \@ifxundefined [1]{%
 \@ifx{#1\undefined}
}%
\providecommand \@ifnum [1]{%
 \ifnum #1\expandafter \@firstoftwo
 \else \expandafter \@secondoftwo
 \fi
}%
\providecommand \@ifx [1]{%
 \ifx #1\expandafter \@firstoftwo
 \else \expandafter \@secondoftwo
 \fi
}%
\providecommand \natexlab [1]{#1}%
\providecommand \enquote  [1]{``#1''}%
\providecommand \bibnamefont  [1]{#1}%
\providecommand \bibfnamefont [1]{#1}%
\providecommand \citenamefont [1]{#1}%
\providecommand \href@noop [0]{\@secondoftwo}%
\providecommand \href [0]{\begingroup \@sanitize@url \@href}%
\providecommand \@href[1]{\@@startlink{#1}\@@href}%
\providecommand \@@href[1]{\endgroup#1\@@endlink}%
\providecommand \@sanitize@url [0]{\catcode `\\12\catcode `\$12\catcode
  `\&12\catcode `\#12\catcode `\^12\catcode `\_12\catcode `\%12\relax}%
\providecommand \@@startlink[1]{}%
\providecommand \@@endlink[0]{}%
\providecommand \url  [0]{\begingroup\@sanitize@url \@url }%
\providecommand \@url [1]{\endgroup\@href {#1}{\urlprefix }}%
\providecommand \urlprefix  [0]{URL }%
\providecommand \Eprint [0]{\href }%
\providecommand \doibase [0]{http://dx.doi.org/}%
\providecommand \selectlanguage [0]{\@gobble}%
\providecommand \bibinfo  [0]{\@secondoftwo}%
\providecommand \bibfield  [0]{\@secondoftwo}%
\providecommand \translation [1]{[#1]}%
\providecommand \BibitemOpen [0]{}%
\providecommand \bibitemStop [0]{}%
\providecommand \bibitemNoStop [0]{.\EOS\space}%
\providecommand \EOS [0]{\spacefactor3000\relax}%
\providecommand \BibitemShut  [1]{\csname bibitem#1\endcsname}%
\let\auto@bib@innerbib\@empty
\bibitem [{\citenamefont {Rong}\ \emph {et~al.}(2004)\citenamefont {Rong},
  \citenamefont {Liu}, \citenamefont {Nicolaescu}, \citenamefont {Paniccia},
  \citenamefont {Cohen},\ and\ \citenamefont {Hak}}]{rong2004}%
  \BibitemOpen
  \bibfield  {author} {\bibinfo {author} {\bibfnamefont {H.}~\bibnamefont
  {Rong}}, \bibinfo {author} {\bibfnamefont {A.}~\bibnamefont {Liu}}, \bibinfo
  {author} {\bibfnamefont {R.}~\bibnamefont {Nicolaescu}}, \bibinfo {author}
  {\bibfnamefont {M.}~\bibnamefont {Paniccia}}, \bibinfo {author}
  {\bibfnamefont {O.}~\bibnamefont {Cohen}}, \ and\ \bibinfo {author}
  {\bibfnamefont {D.}~\bibnamefont {Hak}},\ }\href@noop {} {\bibfield
  {journal} {\bibinfo  {journal} {Appl. Phys. Lett.}\ }\textbf {\bibinfo
  {volume} {85}} (\bibinfo {year} {2004})}\BibitemShut {NoStop}%
\bibitem [{\citenamefont {Rong}\ \emph {et~al.}(2005)\citenamefont {Rong},
  \citenamefont {Jones}, \citenamefont {Liu}, \citenamefont {Cohen},
  \citenamefont {Hak}, \citenamefont {Fang},\ and\ \citenamefont
  {Paniccia}}]{rong2005}%
  \BibitemOpen
  \bibfield  {author} {\bibinfo {author} {\bibfnamefont {H.}~\bibnamefont
  {Rong}}, \bibinfo {author} {\bibfnamefont {R.}~\bibnamefont {Jones}},
  \bibinfo {author} {\bibfnamefont {A.}~\bibnamefont {Liu}}, \bibinfo {author}
  {\bibfnamefont {O.}~\bibnamefont {Cohen}}, \bibinfo {author} {\bibfnamefont
  {D.}~\bibnamefont {Hak}}, \bibinfo {author} {\bibfnamefont {A.}~\bibnamefont
  {Fang}}, \ and\ \bibinfo {author} {\bibfnamefont {M.}~\bibnamefont
  {Paniccia}},\ }\href {\doibase 10.1038/nature03346} {\bibfield  {journal}
  {\bibinfo  {journal} {Nature}\ }\textbf {\bibinfo {volume} {433}},\ \bibinfo
  {pages} {725} (\bibinfo {year} {2005})}\BibitemShut {NoStop}%
\bibitem [{\citenamefont {Jalali}\ \emph {et~al.}(2006)\citenamefont {Jalali},
  \citenamefont {Raghunathan}, \citenamefont {Shori}, \citenamefont {Fathpour},
  \citenamefont {Dimitropoulos},\ and\ \citenamefont {Stafsudd}}]{jalali2006}%
  \BibitemOpen
  \bibfield  {author} {\bibinfo {author} {\bibfnamefont {B.}~\bibnamefont
  {Jalali}}, \bibinfo {author} {\bibfnamefont {V.}~\bibnamefont {Raghunathan}},
  \bibinfo {author} {\bibfnamefont {R.}~\bibnamefont {Shori}}, \bibinfo
  {author} {\bibfnamefont {S.}~\bibnamefont {Fathpour}}, \bibinfo {author}
  {\bibfnamefont {D.}~\bibnamefont {Dimitropoulos}}, \ and\ \bibinfo {author}
  {\bibfnamefont {O.}~\bibnamefont {Stafsudd}},\ }\href {\doibase
  10.1109/JSTQE.2006.885340} {\bibfield  {journal} {\bibinfo  {journal} {IEEE
  J. Sel. Top. Quantum Electron.}\ }\textbf {\bibinfo {volume} {12}},\ \bibinfo
  {pages} {1618} (\bibinfo {year} {2006})}\BibitemShut {NoStop}%
\bibitem [{\citenamefont {Foster}\ \emph {et~al.}(2006)\citenamefont {Foster},
  \citenamefont {Turner}, \citenamefont {Sharping}, \citenamefont {Schmidt},
  \citenamefont {Lipson},\ and\ \citenamefont {Gaeta}}]{foster2006}%
  \BibitemOpen
  \bibfield  {author} {\bibinfo {author} {\bibfnamefont {M.~A.}\ \bibnamefont
  {Foster}}, \bibinfo {author} {\bibfnamefont {A.~C.}\ \bibnamefont {Turner}},
  \bibinfo {author} {\bibfnamefont {J.~E.}\ \bibnamefont {Sharping}}, \bibinfo
  {author} {\bibfnamefont {B.~S.}\ \bibnamefont {Schmidt}}, \bibinfo {author}
  {\bibfnamefont {M.}~\bibnamefont {Lipson}}, \ and\ \bibinfo {author}
  {\bibfnamefont {A.~L.}\ \bibnamefont {Gaeta}},\ }\href {\doibase
  10.1038/nature04932} {\bibfield  {journal} {\bibinfo  {journal} {Nature}\
  }\textbf {\bibinfo {volume} {441}},\ \bibinfo {pages} {960} (\bibinfo {year}
  {2006})}\BibitemShut {NoStop}%
\bibitem [{\citenamefont {Rakich}\ \emph {et~al.}(2012)\citenamefont {Rakich},
  \citenamefont {Reinke}, \citenamefont {Camacho}, \citenamefont {Davids},\
  and\ \citenamefont {Wang}}]{rakichprx}%
  \BibitemOpen
  \bibfield  {author} {\bibinfo {author} {\bibfnamefont {P.~T.}\ \bibnamefont
  {Rakich}}, \bibinfo {author} {\bibfnamefont {C.}~\bibnamefont {Reinke}},
  \bibinfo {author} {\bibfnamefont {R.}~\bibnamefont {Camacho}}, \bibinfo
  {author} {\bibfnamefont {P.}~\bibnamefont {Davids}}, \ and\ \bibinfo {author}
  {\bibfnamefont {Z.}~\bibnamefont {Wang}},\ }\href {\doibase
  10.1103/PhysRevX.2.011008} {\bibfield  {journal} {\bibinfo  {journal} {Phys.
  Rev. X}\ }\textbf {\bibinfo {volume} {2}},\ \bibinfo {pages} {011008}
  (\bibinfo {year} {2012})}\BibitemShut {NoStop}%
\bibitem [{\citenamefont {Shin}\ \emph {et~al.}(2013)\citenamefont {Shin},
  \citenamefont {Qiu}, \citenamefont {Jarecki}, \citenamefont {Cox},
  \citenamefont {Olsson}, \citenamefont {Starbuck}, \citenamefont {Wang},\ and\
  \citenamefont {Rakich}}]{shinnatcomm}%
  \BibitemOpen
  \bibfield  {author} {\bibinfo {author} {\bibfnamefont {H.}~\bibnamefont
  {Shin}}, \bibinfo {author} {\bibfnamefont {W.}~\bibnamefont {Qiu}}, \bibinfo
  {author} {\bibfnamefont {R.}~\bibnamefont {Jarecki}}, \bibinfo {author}
  {\bibfnamefont {J.~A.}\ \bibnamefont {Cox}}, \bibinfo {author} {\bibfnamefont
  {R.~H.}\ \bibnamefont {Olsson}}, \bibinfo {author} {\bibfnamefont
  {A.}~\bibnamefont {Starbuck}}, \bibinfo {author} {\bibfnamefont
  {Z.}~\bibnamefont {Wang}}, \ and\ \bibinfo {author} {\bibfnamefont {P.~T.}\
  \bibnamefont {Rakich}},\ }\href {http://dx.doi.org/10.1038/ncomms2943}
  {\bibfield  {journal} {\bibinfo  {journal} {Nat. Commun.}\ }\textbf {\bibinfo
  {volume} {4}} (\bibinfo {year} {2013})}\BibitemShut {NoStop}%
\bibitem [{\citenamefont {Van~Laer}\ \emph
  {et~al.}(2015{\natexlab{a}})\citenamefont {Van~Laer}, \citenamefont {Kuyken},
  \citenamefont {Van~Thourhout},\ and\ \citenamefont {Baets}}]{roel}%
  \BibitemOpen
  \bibfield  {author} {\bibinfo {author} {\bibfnamefont {R.}~\bibnamefont
  {Van~Laer}}, \bibinfo {author} {\bibfnamefont {B.}~\bibnamefont {Kuyken}},
  \bibinfo {author} {\bibfnamefont {D.}~\bibnamefont {Van~Thourhout}}, \ and\
  \bibinfo {author} {\bibfnamefont {R.}~\bibnamefont {Baets}},\ }\href
  {http://dx.doi.org/10.1038/nphoton.2015.11} {\bibfield  {journal} {\bibinfo
  {journal} {Nat. Photon.}\ }\textbf {\bibinfo {volume} {9}},\ \bibinfo {pages}
  {199} (\bibinfo {year} {2015}{\natexlab{a}})}\BibitemShut {NoStop}%
\bibitem [{\citenamefont {Vidal}\ \emph {et~al.}(2007)\citenamefont {Vidal},
  \citenamefont {Piqueras},\ and\ \citenamefont {Mart\'{i}}}]{Vidal07}%
  \BibitemOpen
  \bibfield  {author} {\bibinfo {author} {\bibfnamefont {B.}~\bibnamefont
  {Vidal}}, \bibinfo {author} {\bibfnamefont {M.~A.}\ \bibnamefont {Piqueras}},
  \ and\ \bibinfo {author} {\bibfnamefont {J.}~\bibnamefont {Mart\'{i}}},\
  }\href {\doibase 10.1364/OL.32.000023} {\bibfield  {journal} {\bibinfo
  {journal} {Opt. Lett.}\ }\textbf {\bibinfo {volume} {32}},\ \bibinfo {pages}
  {23} (\bibinfo {year} {2007})}\BibitemShut {NoStop}%
\bibitem [{\citenamefont {Li}\ \emph {et~al.}(2008)\citenamefont {Li},
  \citenamefont {Pernice}, \citenamefont {Xiong}, \citenamefont {Baehr-Jones},
  \citenamefont {Hochberg},\ and\ \citenamefont {Tang}}]{Li2008}%
  \BibitemOpen
  \bibfield  {author} {\bibinfo {author} {\bibfnamefont {M.}~\bibnamefont
  {Li}}, \bibinfo {author} {\bibfnamefont {W.~H.~P.}\ \bibnamefont {Pernice}},
  \bibinfo {author} {\bibfnamefont {C.}~\bibnamefont {Xiong}}, \bibinfo
  {author} {\bibfnamefont {T.}~\bibnamefont {Baehr-Jones}}, \bibinfo {author}
  {\bibfnamefont {M.}~\bibnamefont {Hochberg}}, \ and\ \bibinfo {author}
  {\bibfnamefont {H.~X.}\ \bibnamefont {Tang}},\ }\href {\doibase
  10.1038/nature07545} {\bibfield  {journal} {\bibinfo  {journal} {Nature}\
  }\textbf {\bibinfo {volume} {456}},\ \bibinfo {pages} {480} (\bibinfo {year}
  {2008})}\BibitemShut {NoStop}%
\bibitem [{\citenamefont {Zhang}\ and\ \citenamefont
  {Minasian}(2012)}]{zhang12}%
  \BibitemOpen
  \bibfield  {author} {\bibinfo {author} {\bibfnamefont {W.}~\bibnamefont
  {Zhang}}\ and\ \bibinfo {author} {\bibfnamefont {R.}~\bibnamefont
  {Minasian}},\ }\href {\doibase 10.1109/LPT.2012.2198638} {\bibfield
  {journal} {\bibinfo  {journal} {IEEE Photon. Technol. Lett.}\ }\textbf
  {\bibinfo {volume} {24}},\ \bibinfo {pages} {1182} (\bibinfo {year}
  {2012})}\BibitemShut {NoStop}%
\bibitem [{\citenamefont {Pant}\ \emph {et~al.}(2014)\citenamefont {Pant},
  \citenamefont {Marpaung}, \citenamefont {Kabakova}, \citenamefont {Morrison},
  \citenamefont {Poulton},\ and\ \citenamefont {Eggleton}}]{pant2014}%
  \BibitemOpen
  \bibfield  {author} {\bibinfo {author} {\bibfnamefont {R.}~\bibnamefont
  {Pant}}, \bibinfo {author} {\bibfnamefont {D.}~\bibnamefont {Marpaung}},
  \bibinfo {author} {\bibfnamefont {I.~V.}\ \bibnamefont {Kabakova}}, \bibinfo
  {author} {\bibfnamefont {B.}~\bibnamefont {Morrison}}, \bibinfo {author}
  {\bibfnamefont {C.~G.}\ \bibnamefont {Poulton}}, \ and\ \bibinfo {author}
  {\bibfnamefont {B.~J.}\ \bibnamefont {Eggleton}},\ }\href {\doibase
  10.1002/lpor.201300154} {\bibfield  {journal} {\bibinfo  {journal} {Laser
  Photon. Rev.}\ }\textbf {\bibinfo {volume} {8}},\ \bibinfo {pages} {653}
  (\bibinfo {year} {2014})}\BibitemShut {NoStop}%
\bibitem [{\citenamefont {Shin}\ \emph {et~al.}(2015)\citenamefont {Shin},
  \citenamefont {Cox}, \citenamefont {Jarecki}, \citenamefont {Starbuck},
  \citenamefont {Wang},\ and\ \citenamefont {Rakich}}]{shinpper}%
  \BibitemOpen
  \bibfield  {author} {\bibinfo {author} {\bibfnamefont {H.}~\bibnamefont
  {Shin}}, \bibinfo {author} {\bibfnamefont {J.~A.}\ \bibnamefont {Cox}},
  \bibinfo {author} {\bibfnamefont {R.}~\bibnamefont {Jarecki}}, \bibinfo
  {author} {\bibfnamefont {A.}~\bibnamefont {Starbuck}}, \bibinfo {author}
  {\bibfnamefont {Z.}~\bibnamefont {Wang}}, \ and\ \bibinfo {author}
  {\bibfnamefont {P.~T.}\ \bibnamefont {Rakich}},\ }\href@noop {} {\bibfield
  {journal} {\bibinfo  {journal} {Nat. Commun.}\ }\textbf {\bibinfo {volume}
  {6}} (\bibinfo {year} {2015})}\BibitemShut {NoStop}%
\bibitem [{\citenamefont {Li}\ \emph {et~al.}(2013)\citenamefont {Li},
  \citenamefont {Lee},\ and\ \citenamefont {Vahala}}]{Li2013}%
  \BibitemOpen
  \bibfield  {author} {\bibinfo {author} {\bibfnamefont {J.}~\bibnamefont
  {Li}}, \bibinfo {author} {\bibfnamefont {H.}~\bibnamefont {Lee}}, \ and\
  \bibinfo {author} {\bibfnamefont {K.~J.}\ \bibnamefont {Vahala}},\ }\href
  {http://dx.doi.org/10.1038/ncomms3097} {\bibfield  {journal} {\bibinfo
  {journal} {Nat. Commun.}\ }\textbf {\bibinfo {volume} {4}} (\bibinfo {year}
  {2013})}\BibitemShut {NoStop}%
\bibitem [{\citenamefont {Marpaung}\ \emph {et~al.}(2015)\citenamefont
  {Marpaung}, \citenamefont {Morrison}, \citenamefont {Pagani}, \citenamefont
  {Pant}, \citenamefont {Choi}, \citenamefont {Luther-Davies}, \citenamefont
  {Madden},\ and\ \citenamefont {Eggleton}}]{marpaung}%
  \BibitemOpen
  \bibfield  {author} {\bibinfo {author} {\bibfnamefont {D.}~\bibnamefont
  {Marpaung}}, \bibinfo {author} {\bibfnamefont {B.}~\bibnamefont {Morrison}},
  \bibinfo {author} {\bibfnamefont {M.}~\bibnamefont {Pagani}}, \bibinfo
  {author} {\bibfnamefont {R.}~\bibnamefont {Pant}}, \bibinfo {author}
  {\bibfnamefont {D.-Y.}\ \bibnamefont {Choi}}, \bibinfo {author}
  {\bibfnamefont {B.}~\bibnamefont {Luther-Davies}}, \bibinfo {author}
  {\bibfnamefont {S.~J.}\ \bibnamefont {Madden}}, \ and\ \bibinfo {author}
  {\bibfnamefont {B.~J.}\ \bibnamefont {Eggleton}},\ }\href {\doibase
  10.1364/OPTICA.2.000076} {\bibfield  {journal} {\bibinfo  {journal} {Optica}\
  }\textbf {\bibinfo {volume} {2}},\ \bibinfo {pages} {76} (\bibinfo {year}
  {2015})}\BibitemShut {NoStop}%
\bibitem [{\citenamefont {Kang}\ \emph {et~al.}(2009)\citenamefont {Kang},
  \citenamefont {Nazarkin}, \citenamefont {Brenn},\ and\ \citenamefont
  {Russell}}]{kang}%
  \BibitemOpen
  \bibfield  {author} {\bibinfo {author} {\bibfnamefont {M.~S.}\ \bibnamefont
  {Kang}}, \bibinfo {author} {\bibfnamefont {A.}~\bibnamefont {Nazarkin}},
  \bibinfo {author} {\bibfnamefont {A.}~\bibnamefont {Brenn}}, \ and\ \bibinfo
  {author} {\bibfnamefont {P.~S.~J.}\ \bibnamefont {Russell}},\ }\href
  {\doibase 10.1038/nphys1217} {\bibfield  {journal} {\bibinfo  {journal} {Nat.
  Phys.}\ }\textbf {\bibinfo {volume} {5}},\ \bibinfo {pages} {276} (\bibinfo
  {year} {2009})}\BibitemShut {NoStop}%
\bibitem [{\citenamefont {Braje}\ \emph {et~al.}(2009)\citenamefont {Braje},
  \citenamefont {Hollberg},\ and\ \citenamefont {Diddams}}]{braje09}%
  \BibitemOpen
  \bibfield  {author} {\bibinfo {author} {\bibfnamefont {D.}~\bibnamefont
  {Braje}}, \bibinfo {author} {\bibfnamefont {L.}~\bibnamefont {Hollberg}}, \
  and\ \bibinfo {author} {\bibfnamefont {S.}~\bibnamefont {Diddams}},\ }\href
  {\doibase 10.1103/PhysRevLett.102.193902} {\bibfield  {journal} {\bibinfo
  {journal} {Phys. Rev. Lett.}\ }\textbf {\bibinfo {volume} {102}},\ \bibinfo
  {pages} {193902} (\bibinfo {year} {2009})}\BibitemShut {NoStop}%
\bibitem [{\citenamefont {Olsson}\ and\ \citenamefont {van~der
  Ziel}(1986)}]{olsson}%
  \BibitemOpen
  \bibfield  {author} {\bibinfo {author} {\bibfnamefont {N.}~\bibnamefont
  {Olsson}}\ and\ \bibinfo {author} {\bibfnamefont {J.}~\bibnamefont {van~der
  Ziel}},\ }\href {\doibase 10.1049/el:19860332} {\bibfield  {journal}
  {\bibinfo  {journal} {Electron. Lett.}\ }\textbf {\bibinfo {volume} {22}},\
  \bibinfo {pages} {488} (\bibinfo {year} {1986})}\BibitemShut {NoStop}%
\bibitem [{\citenamefont {Abedin}\ \emph {et~al.}(2012)\citenamefont {Abedin},
  \citenamefont {Westbrook}, \citenamefont {Nicholson}, \citenamefont {Porque},
  \citenamefont {Kremp},\ and\ \citenamefont {Liu}}]{Abedin12}%
  \BibitemOpen
  \bibfield  {author} {\bibinfo {author} {\bibfnamefont {K.~S.}\ \bibnamefont
  {Abedin}}, \bibinfo {author} {\bibfnamefont {P.~S.}\ \bibnamefont
  {Westbrook}}, \bibinfo {author} {\bibfnamefont {J.~W.}\ \bibnamefont
  {Nicholson}}, \bibinfo {author} {\bibfnamefont {J.}~\bibnamefont {Porque}},
  \bibinfo {author} {\bibfnamefont {T.}~\bibnamefont {Kremp}}, \ and\ \bibinfo
  {author} {\bibfnamefont {X.}~\bibnamefont {Liu}},\ }\href {\doibase
  10.1364/OL.37.000605} {\bibfield  {journal} {\bibinfo  {journal} {Opt.
  Lett.}\ }\textbf {\bibinfo {volume} {37}},\ \bibinfo {pages} {605} (\bibinfo
  {year} {2012})}\BibitemShut {NoStop}%
\bibitem [{\citenamefont {Li}\ \emph {et~al.}(2014)\citenamefont {Li},
  \citenamefont {Lee},\ and\ \citenamefont {Vahala}}]{Li14}%
  \BibitemOpen
  \bibfield  {author} {\bibinfo {author} {\bibfnamefont {J.}~\bibnamefont
  {Li}}, \bibinfo {author} {\bibfnamefont {H.}~\bibnamefont {Lee}}, \ and\
  \bibinfo {author} {\bibfnamefont {K.~J.}\ \bibnamefont {Vahala}},\ }\href
  {\doibase 10.1364/OL.39.000287} {\bibfield  {journal} {\bibinfo  {journal}
  {Opt. Lett.}\ }\textbf {\bibinfo {volume} {39}},\ \bibinfo {pages} {287}
  (\bibinfo {year} {2014})}\BibitemShut {NoStop}%
\bibitem [{\citenamefont {Gao}\ \emph {et~al.}(2013)\citenamefont {Gao},
  \citenamefont {Pant}, \citenamefont {Li}, \citenamefont {Poulton},
  \citenamefont {Choi}, \citenamefont {Madden}, \citenamefont {Luther-Davies},\
  and\ \citenamefont {Eggleton}}]{Gao13}%
  \BibitemOpen
  \bibfield  {author} {\bibinfo {author} {\bibfnamefont {F.}~\bibnamefont
  {Gao}}, \bibinfo {author} {\bibfnamefont {R.}~\bibnamefont {Pant}}, \bibinfo
  {author} {\bibfnamefont {E.}~\bibnamefont {Li}}, \bibinfo {author}
  {\bibfnamefont {C.~G.}\ \bibnamefont {Poulton}}, \bibinfo {author}
  {\bibfnamefont {D.-Y.}\ \bibnamefont {Choi}}, \bibinfo {author}
  {\bibfnamefont {S.~J.}\ \bibnamefont {Madden}}, \bibinfo {author}
  {\bibfnamefont {B.}~\bibnamefont {Luther-Davies}}, \ and\ \bibinfo {author}
  {\bibfnamefont {B.~J.}\ \bibnamefont {Eggleton}},\ }\href {\doibase
  10.1364/OE.21.008605} {\bibfield  {journal} {\bibinfo  {journal} {Opt.
  Express}\ }\textbf {\bibinfo {volume} {21}},\ \bibinfo {pages} {8605}
  (\bibinfo {year} {2013})}\BibitemShut {NoStop}%
\bibitem [{\citenamefont {Yu}\ and\ \citenamefont {Fan}(2009)}]{Yu2009}%
  \BibitemOpen
  \bibfield  {author} {\bibinfo {author} {\bibfnamefont {Z.}~\bibnamefont
  {Yu}}\ and\ \bibinfo {author} {\bibfnamefont {S.}~\bibnamefont {Fan}},\
  }\href {\doibase 10.1038/nphoton.2008.273} {\bibfield  {journal} {\bibinfo
  {journal} {Nat. Photon.}\ }\textbf {\bibinfo {volume} {3}},\ \bibinfo {pages}
  {91} (\bibinfo {year} {2009})}\BibitemShut {NoStop}%
\bibitem [{\citenamefont {Kang}\ \emph {et~al.}(2011)\citenamefont {Kang},
  \citenamefont {Butsch},\ and\ \citenamefont {Russell}}]{Kang2011}%
  \BibitemOpen
  \bibfield  {author} {\bibinfo {author} {\bibfnamefont {M.~S.}\ \bibnamefont
  {Kang}}, \bibinfo {author} {\bibfnamefont {A.}~\bibnamefont {Butsch}}, \ and\
  \bibinfo {author} {\bibfnamefont {P.~S.~J.}\ \bibnamefont {Russell}},\ }\href
  {\doibase 10.1038/nphoton.2011.180} {\bibfield  {journal} {\bibinfo
  {journal} {Nat Photon}\ }\textbf {\bibinfo {volume} {5}},\ \bibinfo {pages}
  {549} (\bibinfo {year} {2011})}\BibitemShut {NoStop}%
\bibitem [{\citenamefont {Huang}\ and\ \citenamefont {Fan}(2011)}]{Huang11}%
  \BibitemOpen
  \bibfield  {author} {\bibinfo {author} {\bibfnamefont {X.}~\bibnamefont
  {Huang}}\ and\ \bibinfo {author} {\bibfnamefont {S.}~\bibnamefont {Fan}},\
  }\href {\doibase 10.1109/JLT.2011.2158886} {\bibfield  {journal} {\bibinfo
  {journal} {J. Lightwave Technol.}\ }\textbf {\bibinfo {volume} {29}},\
  \bibinfo {pages} {2267} (\bibinfo {year} {2011})}\BibitemShut {NoStop}%
\bibitem [{\citenamefont {Kim}\ \emph {et~al.}(2015)\citenamefont {Kim},
  \citenamefont {Kuzyk}, \citenamefont {Han}, \citenamefont {Wang},\ and\
  \citenamefont {Bahl}}]{kim2015}%
  \BibitemOpen
  \bibfield  {author} {\bibinfo {author} {\bibfnamefont {J.}~\bibnamefont
  {Kim}}, \bibinfo {author} {\bibfnamefont {M.~C.}\ \bibnamefont {Kuzyk}},
  \bibinfo {author} {\bibfnamefont {K.}~\bibnamefont {Han}}, \bibinfo {author}
  {\bibfnamefont {H.}~\bibnamefont {Wang}}, \ and\ \bibinfo {author}
  {\bibfnamefont {G.}~\bibnamefont {Bahl}},\ }\href
  {http://dx.doi.org/10.1038/nphys3236} {\bibfield  {journal} {\bibinfo
  {journal} {Nat. Phys.}\ }\textbf {\bibinfo {volume} {11}},\ \bibinfo {pages}
  {275} (\bibinfo {year} {2015})}\BibitemShut {NoStop}%
\bibitem [{\citenamefont {Yao}(1998)}]{Yao98}%
  \BibitemOpen
  \bibfield  {author} {\bibinfo {author} {\bibfnamefont {X.}~\bibnamefont
  {Yao}},\ }\href {\doibase 10.1109/68.651138} {\bibfield  {journal} {\bibinfo
  {journal} {IEEE Photon. Technol. Lett.}\ }\textbf {\bibinfo {volume} {10}},\
  \bibinfo {pages} {138} (\bibinfo {year} {1998})}\BibitemShut {NoStop}%
\bibitem [{\citenamefont {Okawachi}\ \emph {et~al.}(2005)\citenamefont
  {Okawachi}, \citenamefont {Bigelow}, \citenamefont {Sharping}, \citenamefont
  {Zhu}, \citenamefont {Schweinsberg}, \citenamefont {Gauthier}, \citenamefont
  {Boyd},\ and\ \citenamefont {Gaeta}}]{gaeta}%
  \BibitemOpen
  \bibfield  {author} {\bibinfo {author} {\bibfnamefont {Y.}~\bibnamefont
  {Okawachi}}, \bibinfo {author} {\bibfnamefont {M.~S.}\ \bibnamefont
  {Bigelow}}, \bibinfo {author} {\bibfnamefont {J.~E.}\ \bibnamefont
  {Sharping}}, \bibinfo {author} {\bibfnamefont {Z.}~\bibnamefont {Zhu}},
  \bibinfo {author} {\bibfnamefont {A.}~\bibnamefont {Schweinsberg}}, \bibinfo
  {author} {\bibfnamefont {D.~J.}\ \bibnamefont {Gauthier}}, \bibinfo {author}
  {\bibfnamefont {R.~W.}\ \bibnamefont {Boyd}}, \ and\ \bibinfo {author}
  {\bibfnamefont {A.~L.}\ \bibnamefont {Gaeta}},\ }\href {\doibase
  10.1103/PhysRevLett.94.153902} {\bibfield  {journal} {\bibinfo  {journal}
  {Phys. Rev. Lett.}\ }\textbf {\bibinfo {volume} {94}},\ \bibinfo {pages}
  {153902} (\bibinfo {year} {2005})}\BibitemShut {NoStop}%
\bibitem [{\citenamefont {Loayssa}\ \emph {et~al.}(2000)\citenamefont
  {Loayssa}, \citenamefont {Benito},\ and\ \citenamefont {Garde}}]{loayssa}%
  \BibitemOpen
  \bibfield  {author} {\bibinfo {author} {\bibfnamefont {A.}~\bibnamefont
  {Loayssa}}, \bibinfo {author} {\bibfnamefont {D.}~\bibnamefont {Benito}}, \
  and\ \bibinfo {author} {\bibfnamefont {M.~J.}\ \bibnamefont {Garde}},\ }\href
  {\doibase 10.1364/OL.25.001234} {\bibfield  {journal} {\bibinfo  {journal}
  {Opt. Lett.}\ }\textbf {\bibinfo {volume} {25}},\ \bibinfo {pages} {1234}
  (\bibinfo {year} {2000})}\BibitemShut {NoStop}%
\bibitem [{\citenamefont {Van~Laer}\ \emph
  {et~al.}(2015{\natexlab{b}})\citenamefont {Van~Laer}, \citenamefont {Bazin},
  \citenamefont {Kuyken}, \citenamefont {Baets},\ and\ \citenamefont
  {Van~Thourhout}}]{van2015net}%
  \BibitemOpen
  \bibfield  {author} {\bibinfo {author} {\bibfnamefont {R.}~\bibnamefont
  {Van~Laer}}, \bibinfo {author} {\bibfnamefont {A.}~\bibnamefont {Bazin}},
  \bibinfo {author} {\bibfnamefont {B.}~\bibnamefont {Kuyken}}, \bibinfo
  {author} {\bibfnamefont {R.}~\bibnamefont {Baets}}, \ and\ \bibinfo {author}
  {\bibfnamefont {D.}~\bibnamefont {Van~Thourhout}},\ }\href@noop {} {\
  (\bibinfo {year} {2015}{\natexlab{b}})},\ \Eprint
  {http://arxiv.org/abs/1508.06318} {arXiv:1508.06318} \BibitemShut {NoStop}%
\bibitem [{\citenamefont {Wolff}\ \emph
  {et~al.}(2015{\natexlab{a}})\citenamefont {Wolff}, \citenamefont {Van~Laer},
  \citenamefont {Steel}, \citenamefont {Eggleton},\ and\ \citenamefont
  {Poulton}}]{wolff2015brillouin}%
  \BibitemOpen
  \bibfield  {author} {\bibinfo {author} {\bibfnamefont {C.}~\bibnamefont
  {Wolff}}, \bibinfo {author} {\bibfnamefont {R.}~\bibnamefont {Van~Laer}},
  \bibinfo {author} {\bibfnamefont {M.~J.}\ \bibnamefont {Steel}}, \bibinfo
  {author} {\bibfnamefont {B.~J.}\ \bibnamefont {Eggleton}}, \ and\ \bibinfo
  {author} {\bibfnamefont {C.~G.}\ \bibnamefont {Poulton}},\ }\href@noop {}
  {\bibfield  {journal} {\bibinfo  {journal} {arXiv preprint arXiv:1510.00079}\
  } (\bibinfo {year} {2015}{\natexlab{a}})}\BibitemShut {NoStop}%
\bibitem [{\citenamefont {Wolff}\ \emph
  {et~al.}(2015{\natexlab{b}})\citenamefont {Wolff}, \citenamefont {Gutsche},
  \citenamefont {Steel}, \citenamefont {Eggleton},\ and\ \citenamefont
  {Poulton}}]{wolffarxiv}%
  \BibitemOpen
  \bibfield  {author} {\bibinfo {author} {\bibfnamefont {C.}~\bibnamefont
  {Wolff}}, \bibinfo {author} {\bibfnamefont {P.}~\bibnamefont {Gutsche}},
  \bibinfo {author} {\bibfnamefont {M.~J.}\ \bibnamefont {Steel}}, \bibinfo
  {author} {\bibfnamefont {B.~J.}\ \bibnamefont {Eggleton}}, \ and\ \bibinfo
  {author} {\bibfnamefont {C.~G.}\ \bibnamefont {Poulton}},\ }\href@noop {} {\
  (\bibinfo {year} {2015}{\natexlab{b}})},\ \Eprint
  {http://arxiv.org/abs/1508.02458} {arXiv:1508.02458} \BibitemShut {NoStop}%
\bibitem [{\citenamefont {Agrawal}(2007)}]{agrawal2007nonlinear}%
  \BibitemOpen
  \bibfield  {author} {\bibinfo {author} {\bibfnamefont {G.~P.}\ \bibnamefont
  {Agrawal}},\ }\href@noop {} {\emph {\bibinfo {title} {Nonlinear fiber
  optics}}}\ (\bibinfo  {publisher} {Academic press},\ \bibinfo {year}
  {2007})\BibitemShut {NoStop}%
\bibitem [{\citenamefont {Barwicz}\ and\ \citenamefont
  {Haus}(2005)}]{barwicz2005three}%
  \BibitemOpen
  \bibfield  {author} {\bibinfo {author} {\bibfnamefont {T.}~\bibnamefont
  {Barwicz}}\ and\ \bibinfo {author} {\bibfnamefont {H.~A.}\ \bibnamefont
  {Haus}},\ }\href@noop {} {\bibfield  {journal} {\bibinfo  {journal} {J.
  Lightwave Technol.}\ }\textbf {\bibinfo {volume} {23}},\ \bibinfo {pages}
  {2719} (\bibinfo {year} {2005})}\BibitemShut {NoStop}%
\bibitem [{\citenamefont {Dong}\ \emph {et~al.}(2010)\citenamefont {Dong},
  \citenamefont {Qian}, \citenamefont {Liao}, \citenamefont {Liang},
  \citenamefont {Kung}, \citenamefont {Feng}, \citenamefont {Shafiiha},
  \citenamefont {Fong}, \citenamefont {Feng}, \citenamefont {Krishnamoorthy}
  \emph {et~al.}}]{dong2010low}%
  \BibitemOpen
  \bibfield  {author} {\bibinfo {author} {\bibfnamefont {P.}~\bibnamefont
  {Dong}}, \bibinfo {author} {\bibfnamefont {W.}~\bibnamefont {Qian}}, \bibinfo
  {author} {\bibfnamefont {S.}~\bibnamefont {Liao}}, \bibinfo {author}
  {\bibfnamefont {H.}~\bibnamefont {Liang}}, \bibinfo {author} {\bibfnamefont
  {C.-C.}\ \bibnamefont {Kung}}, \bibinfo {author} {\bibfnamefont {N.-N.}\
  \bibnamefont {Feng}}, \bibinfo {author} {\bibfnamefont {R.}~\bibnamefont
  {Shafiiha}}, \bibinfo {author} {\bibfnamefont {J.}~\bibnamefont {Fong}},
  \bibinfo {author} {\bibfnamefont {D.}~\bibnamefont {Feng}}, \bibinfo {author}
  {\bibfnamefont {A.~V.}\ \bibnamefont {Krishnamoorthy}},  \emph {et~al.},\
  }\href@noop {} {\bibfield  {journal} {\bibinfo  {journal} {Opt. Express}\
  }\textbf {\bibinfo {volume} {18}},\ \bibinfo {pages} {14474} (\bibinfo {year}
  {2010})}\BibitemShut {NoStop}%
\bibitem [{\citenamefont {Claps}\ \emph {et~al.}(2004)\citenamefont {Claps},
  \citenamefont {Raghunathan}, \citenamefont {Dimitropoulos},\ and\
  \citenamefont {Jalali}}]{clapsraman}%
  \BibitemOpen
  \bibfield  {author} {\bibinfo {author} {\bibfnamefont {R.}~\bibnamefont
  {Claps}}, \bibinfo {author} {\bibfnamefont {V.}~\bibnamefont {Raghunathan}},
  \bibinfo {author} {\bibfnamefont {D.}~\bibnamefont {Dimitropoulos}}, \ and\
  \bibinfo {author} {\bibfnamefont {B.}~\bibnamefont {Jalali}},\ }\href
  {\doibase 10.1364/OPEX.12.002774} {\bibfield  {journal} {\bibinfo  {journal}
  {Opt. Express}\ }\textbf {\bibinfo {volume} {12}},\ \bibinfo {pages} {2774}
  (\bibinfo {year} {2004})}\BibitemShut {NoStop}%
\end{thebibliography}%
\newpage



\section*{Supplementary material}

# Large Brillouin Amplification in Silicon

Eric A. Kittlaus[1], Heedeuk Shin[2], and Peter T. Rakich[1]

[1]*Department of Applied Physics, Yale University, New Haven, CT 06520 USA.*

[2]*Department of Physics, Pohang University of Science and Technology, Pohang 37673, Korea.*

## S1. Two-photon Absorption and Free Carrier Absorption

Due to the nontrivial dynamics of highly-confined optical modes, nonlinear coefficients have effective values different from those estimated using the bulk parameters and traditional definition of the effective mode area [S1]. Using the method of Ref. [S1] and bulk values of $n_2 = 4.5 \times 10^{-18}$ m²W⁻¹ and $\beta_{TPA} = 7.9 \times 10^{-12}$ mW⁻¹ [S2], we calculate $\gamma_k = 110 \pm 17$ m⁻¹W⁻¹ and $\beta_{TPA} = 48 \pm 14$ m⁻¹W⁻¹ for the waveguide Kerr and two-photon absorption coefficients, respectively.

## S2. Nonlinear Loss Model

The propagation loss in the silicon waveguide system is generally captured by the differential equation

$$\frac{dP}{dz} = -\alpha P - \beta P^2 - \gamma P^3 .$$

Here $\alpha$ is the linear loss coefficient and $\beta$ and $\gamma$ are the nonlinear loss coefficients due to two-photon absorption (TPA) and TPA-induced free carrier absorption (FCA), respectively. The guided power at any position $z$, $P(z)$, is defined through the implicit solution to this equation:

$$z - c = \frac{\log\left(\frac{\alpha}{P(z)^2} + \frac{\beta}{P(z)} + \gamma\right)}{2\alpha} + \frac{\beta \tan^{-1}\left(\frac{\beta + 2\gamma P(z)}{\sqrt{4\alpha\gamma - \beta^2}}\right)}{\alpha\sqrt{4\alpha\gamma - \beta^2}} .$$

Here, c is a constant fixed by boundary conditions. $\alpha$ was determined through cutback measurements (Plotted in Figure S1b), and $\beta$ is set to the calculated value of $\beta_{TPA}$. Linear loss is

measured to be $.18 \pm .02$ dB/cm ($\alpha = 4.1 \pm .5$ m$^{-1}$). This model was fit to power-dependent transmission data for the silicon waveguide device using Mathematica to determine $\gamma = 2550 \pm 450$ m$^{-1}$W$^{-2}$. This corresponds to a free carrier lifetime of about 2.2 ns [S4], in excellent agreement with the lifetime estimated by the method of Section S4. Transmission data and a fit to the nonlinear loss model are plotted in Figure S1a.

The effective propagation length in this model is defined as

$$L_{\text{eff}} = \frac{1}{P(0)} \int_0^L P(z)\, dz\,,$$

where $P(z)$ is the total in-device power at position $z$. When only linear loss is present, this reduces to the familiar expression $L_{\text{eff}} = (1 - e^{-\alpha L})/\alpha$.

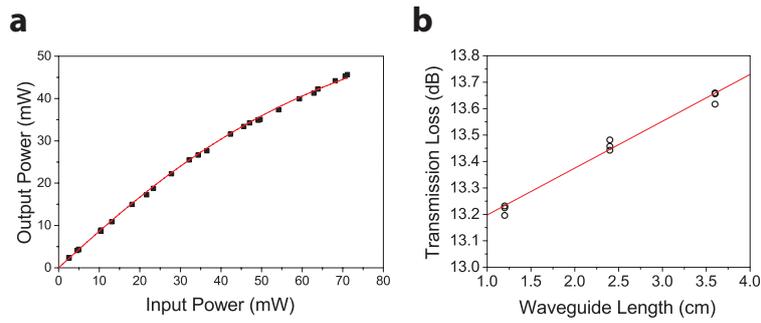

**Figure S1.** Input vs. output power for the device under study showing (a) nonlinear loss as input power is increased. Fit is to the nonlinear loss model described in Section S2. (b) Total transmission loss versus waveguide length for 9 test waveguides at an optical wavelength of 1550 nm, used to determine linear loss.

## S3. Heterodyne Four Wave Mixing Experiment

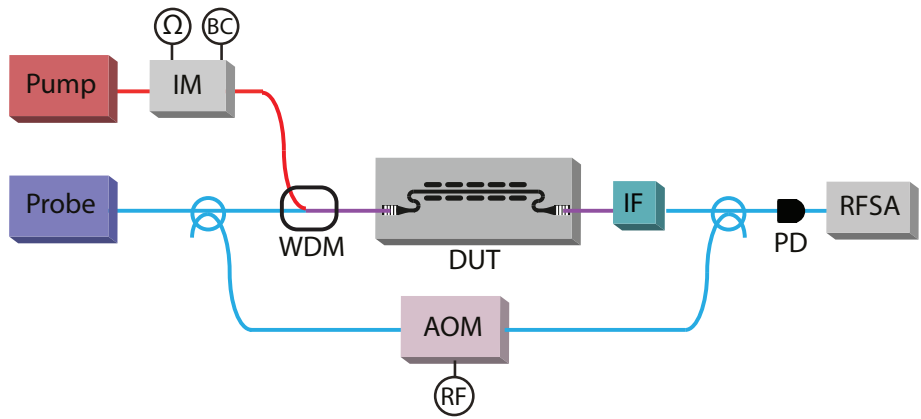

**Figure S2.** Diagram of the Heterodyne four-wave mixing experiment. The Brillouin-active phonon mode is driven by strong pump fields separated by frequency $\Omega$, resulting in nonlinearly-induced phase changes to the probe beam. An acousto-optic modulator provides a reference field for heterodyne detection.

To corroborate direct gain measurements, a heterodyne four-wave mixing (FWM) experiment diagrammed in Figure S2 was performed. This experiment is similar to that performed in Ref. [S3]. The observed lineshape is a result of interference between resonant Brillouin process and the background Kerr nonlinearity, with the normalized fitting function [S3]:

$$\frac{g_{sbs}}{g_0} = \left| e^{i\phi} + \frac{G_B \, L_{sbs}}{L_{tot} \gamma_K^{(3)}} \frac{\frac{\Omega_B}{2Q}}{\Omega_B - \Omega - \frac{i\Omega_B}{2Q}} \right|^2 .$$

Here $\phi$ is a relative phase between the Brillouin and background nonlinearities, $L_{sbs}/L_{tot}$ is the ratio of Brillouin active to total device lengths, and $\gamma_K^{(3)} = 110$ m⁻¹W⁻¹ is the Kerr nonlinear coefficient from section S1. Plots for devices of length 0.5 mm and 29 mm are shown in Figure S3. The nonlinear fit gives $Q = 680$ and $G_B = 1120 \pm 180$ m⁻¹W⁻¹ for the longer device, in excellent agreement with the direct gain measurements (see Figure 2).

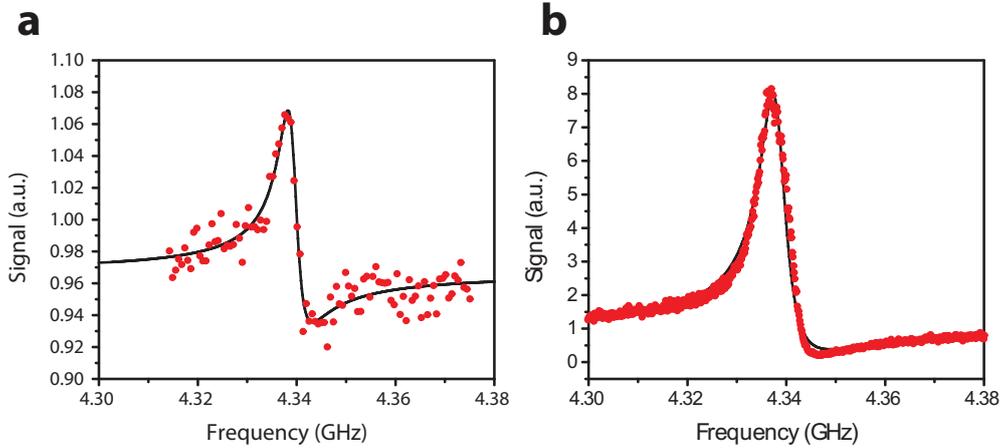

**Figure S3.** Traces (red dots) and fits (black lines) obtained from the heterodyne four wave mixing experiment as the beat frequency of the pump field is swept through a Brillouin resonance for two devices of lengths 0.5 mm (a) and 29 mm (b). These data demonstrate modest broadening of the Brillouin resonance due to device inhomogeneities, discussed in section S7.

## S4. Free Carrier Dynamics for Heterodyne Experiment

At high pump powers, the heterodyne FWM experiment gains an additional background signal component due to fifth-order free carrier nonlinearities. Due to finite free carrier lifetimes, this background component varies with the modulation frequency of the double pump fields and can be used to estimate the system's effective free carrier lifetime. Here we assume we are far detuned from Brillouin resonance, and label the fields as follows: Two bright pump fields $A_1$ and $A_2$ of equal magnitude and separated by frequency $\Omega$. Phase modulation of a probe field $A_4$ generates Stokes and anti-Stokes fields $A_3$ and $A_5$ at relative frequencies of $\pm\Omega$.

Two-photon absorption creates free carriers as the fields propagate, which in turn can induce time-varying refractive index and absorption profiles as follows [S5]:

$$\Delta n = -\frac{e^2 \lambda^2}{8\pi^2 c^2 \epsilon_o n}\left[\frac{\Delta N_e}{m_e} + \frac{\Delta N_h}{m_h}\right] \approx -\frac{e^2 \lambda^2}{8\pi^2 c^2 \epsilon_o n}\left[\frac{1}{m_e} + \frac{1}{m_h}\right] \cdot \Delta N = Q \cdot \Delta N\,,$$

$$\Delta\alpha = \frac{e^3 \lambda^2}{4\pi^2 c^3 \epsilon_o n}\left[\frac{\Delta N_e}{m_e^2 \mu_e} + \frac{\Delta N_h}{m_h^2 \mu_h}\right] \approx \frac{e^3 \lambda^2}{4\pi^2 c^3 \epsilon_o n}\left[\frac{1}{m_e^2 \mu_e} + \frac{1}{m_h^2 \mu_h}\right] \cdot \Delta N = V \cdot \Delta N\,.$$

The free carrier dynamics are governed by the differential equation [S5]

$$\frac{dN}{dt} = -\frac{N}{\tau} + R(t)\,,$$

where $R(t)$ is the free carrier generation rate per unit volume produced by two-photon absorption, which can be expressed as

$$R(t) = \frac{[P_p(t)^2]}{2\hbar\omega} \cdot \frac{\beta_{\text{TPA}}}{A_{\text{eff}}}\,.$$

We consider the free carrier modulation at frequency $\Omega$ since this will affect the fields $A_3$ and $A_5$. The time modulated component of the free carrier density takes the form

$$\Delta N(t) = Re\{\widetilde{N_o}(\Omega)e^{-i\Omega t}\}\,,$$

where

$$\widetilde{N}(\Omega) = \frac{4\gamma_i |A_p|^4}{\hbar\omega A_{\text{eff}}} \cdot \left[\frac{1}{\tau} - i\Omega\right]^{-1}\,,$$

and $A_p$ is the amplitude of either of either driving field (i.e., $p = \{1,2\}$).

Using these results, the coupled amplitude equations for the five fields become:

$$\frac{dA_1}{dz} = i\gamma_k[|A_1|^2 + 2|A_2|^2 + 2|A_4|^2] \cdot A_1\,,$$

$$\frac{dA_2}{dz} = i\gamma_k[|A_2|^2 + 2|A_1|^2 + 2|A_4|^2] \cdot A_2 \,,$$

$$\frac{dA_3}{dz} = i(2\gamma_{\text{FWM}}^*)A_1A_2^*A_4 + \left[i\frac{\omega}{c} \cdot \frac{Q}{2} + \frac{V}{4}\right] \cdot \tilde{N}^*(\Omega) \cdot A_4 \,,$$

$$\frac{dA_4}{dz} = i\gamma_k[|A_4|^2 + 2|A_1|^2 + 2|A_2|^2] \cdot A_4 \,,$$

$$\frac{dA_5}{dz} = i(2\gamma_{\text{FWM}})A_1^*A_2A_4 + \left[i\frac{\omega}{c} \cdot \frac{Q}{2} + \frac{V}{4}\right] \cdot \tilde{N}(\Omega) \cdot A_4 \,.$$

Here $\gamma_k$ is the complex third order nonlinear coefficient, including the effects of Kerr and associated TPA. $\gamma_{\text{FWM}} = Re\{\gamma_k\}$ is the nonlinear coefficient for the four-wave mixing process. These equations lead to a nontrivial free carrier background that depends strongly on pump and probe powers, system parameters, and free carrier lifetime $\tau$.

Simulated results for parameters near estimated experimental conditions are plotted atop experimentally measured data from the heterodyne four-wave mixing experiment (see Section S3) in Figure S4. Good agreement is achieved for both probe fields with an estimated free carrier lifetime of $\tau = 2.3$ ns.

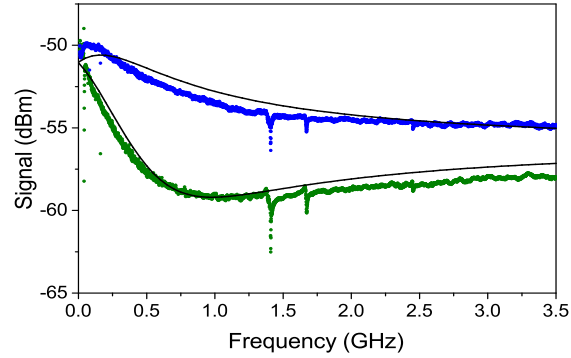

**Figure S4.** Traces for measured anti-Stokes (blue, upper curve) and Stokes (green, lower curve) signals and corresponding theoretical curves as a function of pump modulation frequency. Several Brillouin resonances are also visible. Total input power is around 70 mW.

## S5. Measurement of Anti-stokes Attenuation

In addition to producing amplification, the forward SBS process can also be used as a narrowband filter for light at frequency $\omega_a = \omega_p + \Omega_B$, which can be quantified using the gain experiment of Figure 2a. At the highest tested power, anti-Stokes attenuation reaches over 11.1

dB, with a FWHM (full width at half maximum transmission) of about 7 MHz. Anti-stokes spectra and corresponding peak attenuation are plotted in Figure S5.

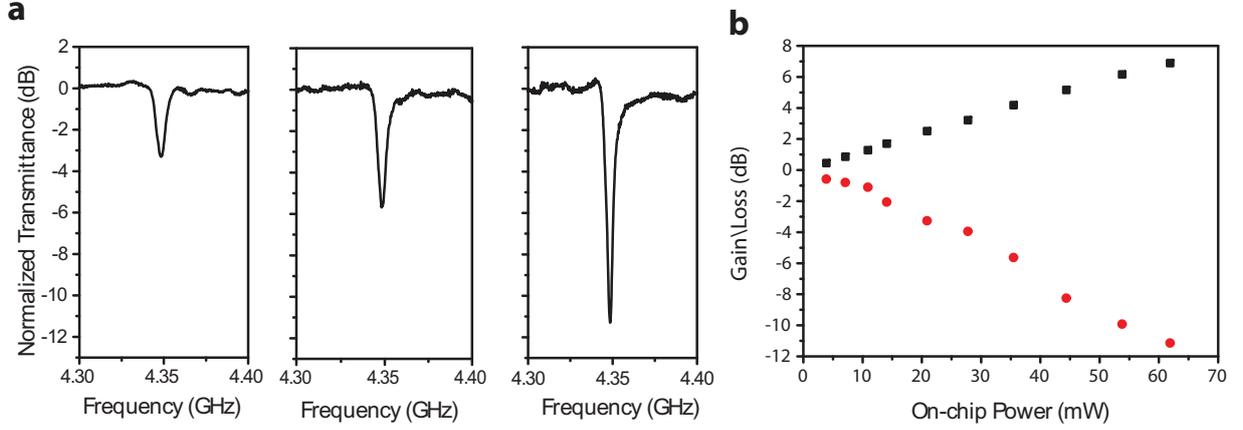

**Figure S5.** Anti-Stokes attenuation of a weak seed beam as a function of pump power, in comparison with Figure 2b-c. (a) Transmittance spectra as pump powers of 21, 36, and 62 mW, respectively. (b) anti-Stokes attenuation grows with increasing pump power. Stokes gain is plotted for comparison.

## S6. Energy Transfer Model

A theoretical model for the energy transfer experiment (see Figure 3) was computed following the coupled amplitude formalism of Ref. [S6]. Nonlinear loss was included according to the model described in Section S2. The coupled amplitude equations for $2N+1$ coupled fields are the set of coupled nonlinear differential equations

$$A'_n(x) = -i\chi\left[\frac{A_{n-1}(x)}{\Omega - \Omega_0 + i\frac{\Gamma}{2}}\sum_{i=-N}^{N} A^*_{i-1}(x)A_i(x) + \frac{A_{n+1}(x)}{\Omega - \Omega_0 - i\frac{\Gamma}{2}}\sum_{i=-N}^{N} A^*_i(x)A_{i-1}(x)\right]$$
$$-\frac{1}{2}\left(\alpha + \beta P(x) + \gamma P^2(x)\right)A_n(x) .$$

The coupling coefficient was calculated through full acoustic and optical mode simulations from the finite-element solver COMSOL using the photoelastic tensor components for silicon ($p_{11}, p_{12}, p_{44}$) = (−0.09, 0.017, −0.051) [S7]. These equations are solved for experimental parameters and coupling rate $\chi$ calculated from a finite element simulation to determine the theoretical model for Figure 3c.

## S7. Sensitivity to Dimensional Variations

Dimensional inhomogeneities along the device length can lead to effective broadening of the Brillouin resonance. The variations in acoustic resonance frequency versus five relevant dimensions was calculated via a finite element simulation and is tabulated below:

| Dimension | Description | Sensitivity to Dimension Error |
|-----------|-------------|-------------------------------|
| a | Total membrane width | $1.4 \times 10^6$ Hz/nm |
| b | Membrane Thickness | $8.3 \times 10^5$ Hz/nm |
| c | Ridge waveguide width | $7.5 \times 10^5$ Hz/nm |
| d | Ridge waveguide height | $7 \times 10^3$ Hz/nm |
| e | Ridge offset from center | $5 \times 10^3$ Hz/nm |

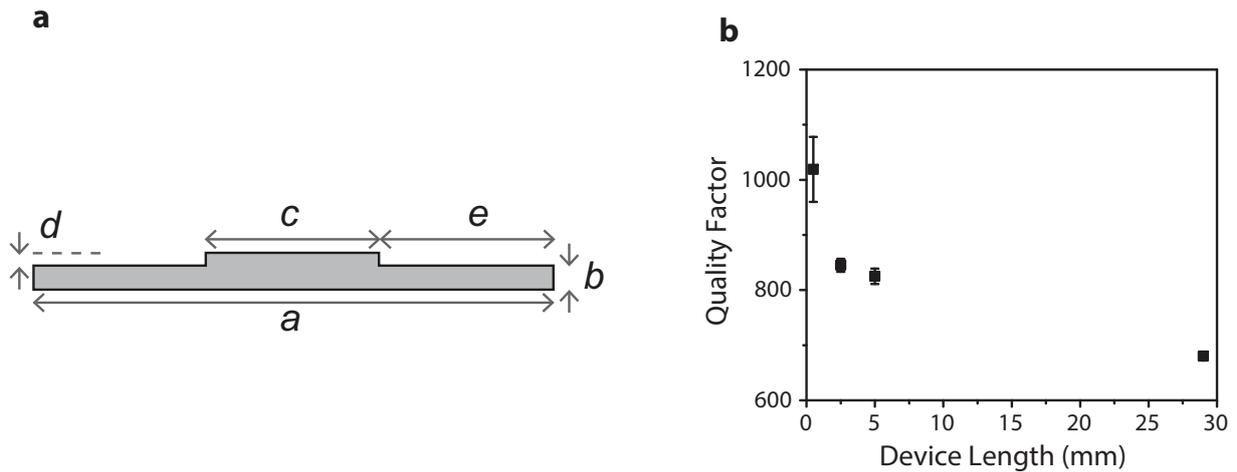

**Figure S6.** (a) Diagram of dimensions specifying device cross section. (b) Effective quality factor measured for devices of different lengths, demonstrating resonance broadening for longer devices.

The largest sensitivity is to total membrane width, though waveguide width and membrane height also play a significant role. These sensitivities are roughly an order of magnitude lower than those for nanowire designs [S8]. Variations in ridge height and offset from the membrane center play almost no role in inhomogeneous broadening of the acoustic resonance. Figure S6b plots measured quality factors for a selection of devices (total active length = 0.5mm, 2.5mm, 5mm, and 29 mm, respectively).

## References

[S1] S. Afshar V and T. M. Monro, Opt. Express **17**, 2298-2318 (2009)

[S2] M. Dinu, F. Quochi, and H. Garcia, Appl. Phys. Lett. **82**, 2954-2956 (2003), DOI:http://dx.doi.org/10.1063/1.1571665

[S3] H. Shin, W. Qiu, R. Jarecki, J. A. Cox, R. H. Olsson, A. Starbuck, Z. Wang, and P. T. Rakich, Nat. Commun. **4** (2013).

[S4] R. Claps, V. Raghunathan, D. Dimitropoulos, and B. Jalali, Opt. Express **12**, 2774-2780 (2004)

[S5] R. Dekker, N. Usechak, M. Först, and A. Driessen, J. Phys. D **40**, R249-R271 (2007).

[S6] M. S. Kang, A. Nazarkin, A. Brenn, and P. S. J. Russell, Nat. Phys. **5**, 276 (2009).

[S7] D. K. Biegelsen, Phys. Rev. Lett., **32**, 1196-1199 (1974).

[S8] R. Van Laer, B. Kuyken, D. Van Thourhout, and R. Baets, Nat. Photon. **9**, 199 (2015).



\end{document}